\documentclass[journal]{IEEEtran}

\usepackage{graphicx}
\ifCLASSINFOpdf
\else
\fi
\usepackage{url}
\usepackage{amsfonts}
\usepackage[noadjust]{cite}

\usepackage{xcolor}

\title{Blockchain-enabled Network Sharing for O-RAN in 5G and Beyond}

\author{Lorenza Giupponi and Francesc Wilhelmi\thanks{The authors are with Centre Tecnol\`ogic de Telecomunicacions de Catalunya (CTTC/CERCA).}}
\begin{document}
\maketitle

\begin{abstract}
The innovation provided by network virtualization in 5G, together with standardization and openness boosted by the Open Radio Access Network (O-RAN) Alliance, has paved the way to a collaborative future in cellular systems, driven by flexible network sharing. Such advents are expected to attract new players like content providers and verticals, increasing competitiveness in the telecom market. However, scalability and trust issues are expected to arise, given the criticality of ownership traceability and resource exchanging in a sharing ecosystem. To address that, we propose integrating blockchain technology for enabling mobile operators and other players to exchange RAN resources (e.g., infrastructure) in the form of virtual network functions (VNF) autonomously and dynamically. Blockchain will provide automation, robustness, trustworthiness, and reliability to mobile networks, thus bringing confidence to open RAN environments. In particular, we define a novel O-RAN-based blockchain-enabled architecture that allows automating RAN sharing procedures through either auction or marketplace-based mechanisms. The potential advantages of the proposed solution are demonstrated through simulation results. The used simulation platform is openly released.
\end{abstract}


\IEEEpeerreviewmaketitle

%%%%%%%%%%%%%%%%%%%%%%%%%%%%
%% INTRODUCTION
%%%%%%%%%%%%%%%%%%%%%%%%%%%%
\section{Introduction}

% 1 MOTIVATION WITH KEY REFERENCES
As of today, 5G is a reality, with 100+ operators worldwide already deploying it. Despite that, no matter the better capacity and latency offered by 5G, an increment in average revenue per user (ARPU) is unclear, and the past trends show that this indicator has been steadily decreasing for over a decade. A remarkable aspect in the economics of 5G is that 70\% of network management costs are concentrated in the radio access network (RAN), where novel requirements (e.g., to pursue the millimeter-wave vision) entail using more spectrum, upgrading equipment, or deploying large-scale small-cells~\cite{webb20185g}. This has generated an increasing interest to cut both capital and operational expenditures (CAPEX and OPEX) and increase automation in the RAN~\cite{ORANall}. An appealing solution for cost effective management of the network has been identified by industry and academia in the concept of network and RAN sharing. This allows operators and service providers to exchange network resources to satisfy the increasing users' demands, without the need for significant investments to enter the market and remain competitive~\cite{samdanis2016network}.

The need for economic sustainability of future networks has paved the way to two main trends in the industry. First, evolving inside 3rd Generation Partnership Project (3GPP) and Next Generation Mobile Networks (NGMN), next-generation self-organizing networks (NG-SON) for 5G and beyond are meant to introduce automation through artificial intelligence (AI) techniques. Second, Network Function Virtualization (NFV) has attracted a lot of interest by allowing to split out dedicated capacity through \textit{network slicing}~\cite{foukas2017network}, and deploy custom virtual network functions (VNFs) to meet specific user demands and minimize network investments.

% 2 MAIN RELATED WORK WITH RESEARCH CHALLENGES
To leverage the potential of AI and network virtualization in next-generation communications, novel architectural transformations are proposed for the RAN through global initiatives like Open-RAN (O-RAN)~\cite{ORANall}, which aims to introduce virtualized network elements, openness, and intelligence to RAN management. O-RAN has been targeted as a strong enabler for RAN sharing since it allows dealing with multi-vendor deployments by dynamically managing and orchestrating both radio and cloud resources~\cite{ORANall}, supported by embedded AI functionalities. The novel architecture proposed by O-RAN generates new opportunities related to the offering, distribution, and execution of VNFs by other operators. Stakeholders can quickly contract and deploy VNFs from catalogs (i.e., marketplaces) to support user equipment (UE) demands. Operators with available resources on their site can openly expose their service conditions and prices so that any other player can pick the most suitable offer. With this approach, competition is open, but quite static, as prices are not adapted to specific RAN users' demands. Moreover, considering the openness of O-RAN sharing environments, novel mechanisms are required to enable dynamic and real-time competitive resource trading securely and reliably.

% 3 GOALS, CONTRIBUTIONS, AND RESULTS
To support and provide trust in these emerging open markets, we propose the introduction of blockchain and smart contracts technologies in O-RAN management, which offer immutable and permanent records, for interested parties to audit. The smart contract is used to describe the RAN user's requirements and allow for SLA enforcement. On top of the automation and the network management efficiency proposed by NG-SON and O-RAN, blockchain removes the need for costly intermediaries (e.g., bank, credit rating agency) and enables unprecedented levels of transparency and trustworthiness, with the potential for high savings. In addition, blockchain reduces the delay required to establish agreements facilitating a real dynamism in RAN sharing.

\begin{figure}[ht!]
\includegraphics[width=\linewidth]{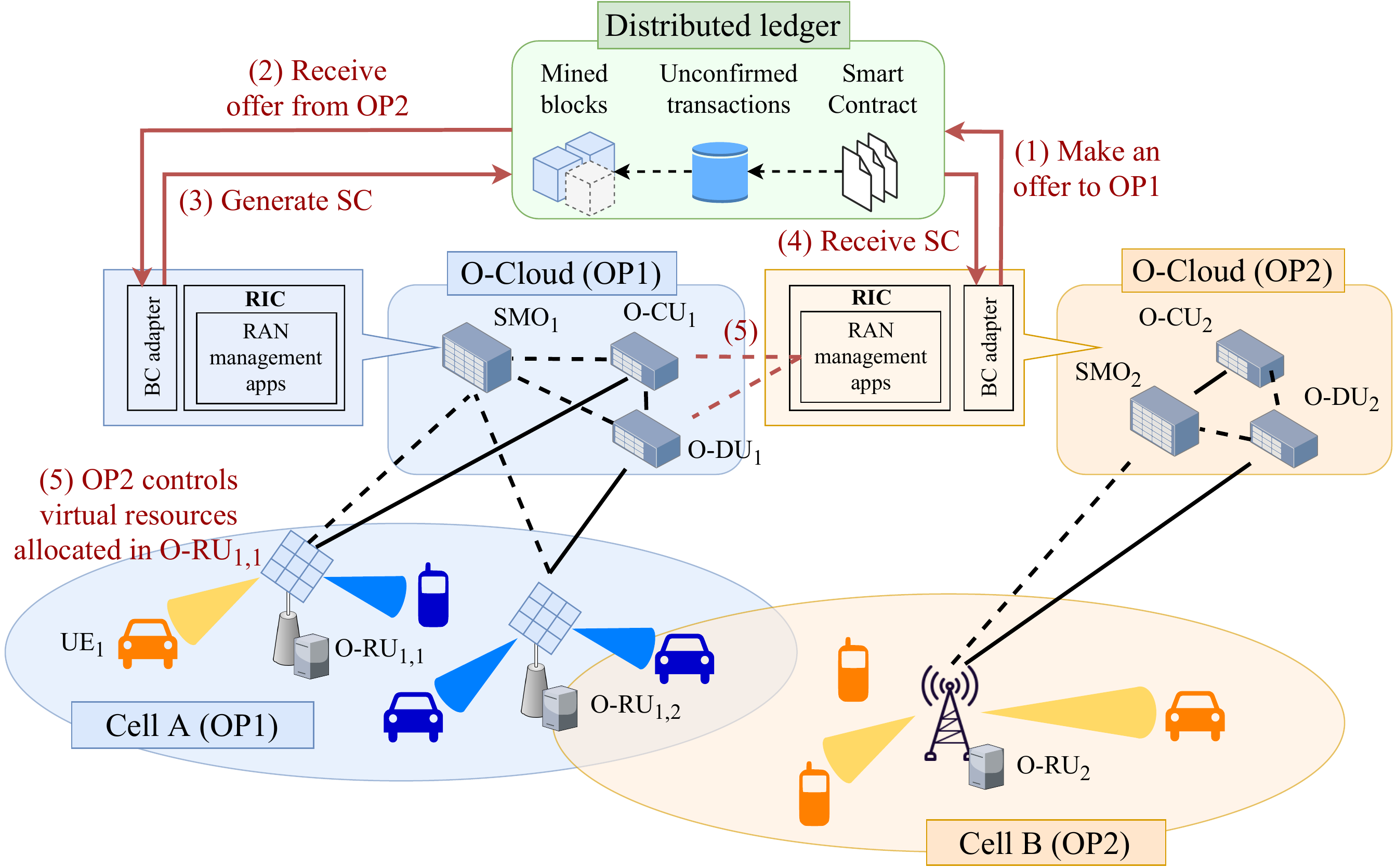}
\caption{Blockchain-enabled RAN sharing ecosystem.}
\label{fig:blockchain_ecosystem}
\end{figure}

Through blockchain-enabled RAN sharing for 5G and beyond, where O-RAN is the baseline architecture, operators dynamically sublease their resources to capitalize on the available infrastructure and allow other operators to increase coverage and capacity. Each operator is allowed to optimally choose between capital investment and resource use continuously, and not only at the time of RAN sharing contract signature, or network deployment. Dynamic resource trading allows for the establishment of new competitive markets, where new stakeholders appear, democratizing and decentralizing the telecom market. The overall concept is depicted in Fig.~\ref{fig:blockchain_ecosystem}, where we introduce two O-RAN-based operators (OP1 and OP2), which agree on leasing certain network resources in the form of VNFs, relying on blockchain and smart contract as technologies enabling trust and automation. More details regarding the purpose and interfaces of O-RAN components are provided next.

The introduction of blockchain in network management has, however, also well-documented limitations (see \cite{FWilhelmi_PIMRC} and references therein). Maintaining a blockchain among a wide number of peers incurs heavy costs in computation, power, energy consumption, and memory usage. Running a blockchain in a mobile network represents a high traffic overhead due to the distribution of blocks across blockchain participants. The additional delay is another aspect that affects the performance, the service perception, and the stability of the blockchain itself. Taking into account these limitations is essential when designing solutions such as the one proposed in this paper (e.g., whether to use a public or a private blockchain, which type of consensus mechanism to apply, how to tune blockchain parameters). The target of this paper is to propose a blockchain-enabled O-RAN-based architecture for beyond 5G and to evaluate the advantages and disadvantages incurred by the introduction of the blockchain in O-RAN management. Evaluation is done in terms of delay and overhead for service establishment as a function of the number of participating operators and blockchain parameters.

%In this context, reverse auctions have been proposed to enable flexible competition among providers~\cite{franco2019brain}, favoring lower prices while meeting users' needs.

%%%%%%%%%%%%%%%%%%%%%%%%%%%%
%% TUTORIAL SECTION ON ORAN
%%%%%%%%%%%%%%%%%%%%%%%%%%%%
\section{Background on O-RAN Architecture}
\label{section:oran}
The O-RAN Alliance was born in February 2018 to continue the evolution of 3GPP RAN architecture in areas like non-public networks, self-organized networks, and integrated access and backhaul. At this time, the O-RAN initiative has received great acceptance in the industry, with over 160 companies supporting it~\cite{garcia2021ran}. The major purpose of O-RAN is to define open interfaces between elements implemented in general-purpose hardware. It is the first standard to enable multi-vendor RAN, and RAN virtualization, thus favoring efficient splits over the protocol stack for network slicing purposes. This flexibility allows to bring in new radio players and gives plenty of opportunities to operators to optimize deployments for specific performance requirements at a much better cost. 

\begin{figure*}[ht!]    
\centering
\includegraphics[width=0.7\textwidth]{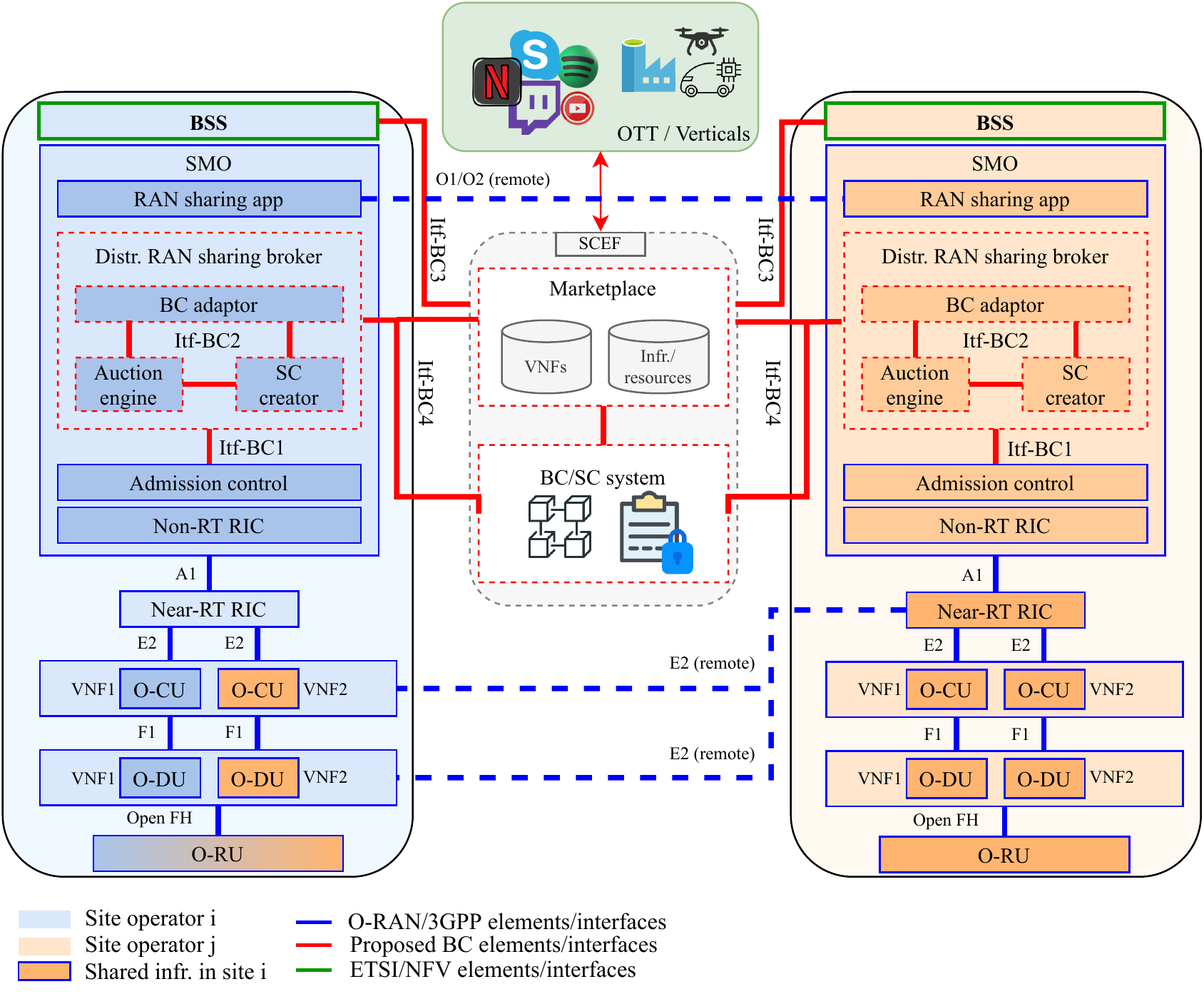}
\caption{Functional blockchain-enabled O-RAN architecture.}
\label{fig:functionalarchitecture}
\end{figure*}

The O-RAN architecture, based on the original 3GPP proposal in TS 38.401, is shown in Fig.~\ref{fig:functionalarchitecture} for two operators sharing a coverage area. Specifically, the figure represents a RAN sharing-compliant O-RAN architecture that allows operators to remotely configure the shared network resources independently from configuration and operating strategies of the hosting operator. This RAN sharing use case is extensible to MVNOs, service providers, and verticals requesting RAN resources. The main O-RAN modules (outer lined in blue), deployed as VNFs or containers, and whose interfaces are defined by O-RAN in~\cite{ORANall}, are as follows: 
\begin{itemize}
    \item \textbf{System management and orchestration (SMO):} The SMO, similar to other architectures based on ETSI NFV's, provides a variety of network management functionalities. In the context of O-RAN and for what concerns the RAN functionalities, it oversees all the orchestration, management, and automation aspects. It includes fault, configuration, accounting, performance, security (FCAPS) support for O-RAN functions, such as installation, configuration, or performance, fault, and file management. In addition, it includes the newly introduced non-real-time (non-RT) Radio Intelligent Controller (RIC) for intelligent RAN optimization. These functionalities are supported by newly introduced interfaces such as A1, O1, and O2.
    \item \textbf{O-RAN centralized unit (O-CU):} It hosts the gNB protocols of Radio Resource Control (RRC), Service Data Adaptation Protocol (SDAP), and Packet Data Convergence Protocol (PDCP), and is connected to the O-DU via F1-C and F1-U interfaces, for control and data planes, respectively. The O-CU controls the operation of multiple O-DU over midhaul interfaces. This split architecture, although not the only possible one, enables different distributions of protocol stack between O-CU and O-DU, depending on the midhaul availability and the network design.
    \item \textbf{O-RAN decentralized unit (O-DU):} It hosts the radio link control (RLC), the medium access control (MAC), and PHY-high layers of the gNB, and is connected to the O-RU via the Open Fronthaul interface. This node includes a subset of the eNB/gNB functions, depending on the functional split option, and its operation is controlled by the O-CU.
   \item \textbf{O-RAN radio unit (O-RU):} It hosts the PHY-low layers of the gNB, and carries out tasks related to RF processing, following the functional split provided by the O-RAN architecture.
    \item \textbf{Radio intelligent controller (RIC):} O-CUs are controlled by the non-RT and the near-RT RIC modules, supporting tasks with more or less than one~second latency, respectively. Through these modules, the {O-RAN} architecture strives the industry to introduce embedded AI in the system. RICs are based on software apps to orchestrate and manage the RAN. Functionalities like mobility management, admission control, and interference management are available as apps on the RIC, which enforces network policies to the radio segment. Non-RT RIC is implemented inside the SMO and includes service and policy management, RAN analytics, and model training for the near-RT RIC. The trained model is managed by the SMO for deployment and received by the near-RT RIC through A1 interface. The near-RT RIC uses the E2 interface to collect information at UE and cell basis and perform self-optimization tasks across the heterogeneous RAN (e.g., formed by macros, massive MIMO, small cells).
\end{itemize}
O-RAN has studied different options for CU/DU/RU deployment, including location in regional clouds, or operator-specific sites. The so-called \textit{Scenario B} is the initial priority of {O-RAN}, where O-CU and O-DU are located together in a cloud, while the O-RU is located in a proprietary cell site.

%%%%%%%%%%%%%%%%%%%%%%%%%%%%
%% TUTORIAL SECTION ON BC AND NETWORK/RAN SHARING
%%%%%%%%%%%%%%%%%%%%%%%%%%%%
\section{A Primer on blockchain-enabled Network/RAN Sharing in cellular networks}
\label{section:related_work}

A blockchain is a type of distributed ledger technology (DLT) compiling transactions in blocks that are sequentially and cryptographically chained one after the other. A peer-to-peer (P2P) network is in charge of maintaining the transactions' record on the blockchain. The miners organize the unconfirmed transactions into blocks and append them to the blockchain after achieving consensus. The consensus is executed by all the blockchain participants to validate new blocks without the need to rely on a central trusted authority. Blockchain was firstly introduced with the cryptocurrency Bitcoin. Successively, with the irruption of smart contracts, i.e., computer programs that run on top of the blockchain and self-execute the terms of a contract when specific conditions are met, blockchain's applications appeared in multiple domains.%~\cite{cai2018decentralized}. 

The advent of blockchain raises issues related to memory usage, overhead, power consumption, or latency, and require to be optimized by the blockchain protocol itself. Besides, interoperability issues, both at a technical level (how interfaces talk to each other) and at a semantic level (how information is understood by the parties involved), need to be addressed by standardization activities~\cite{konig2020comparing}. In this regard, important organizations such as the International Telecommunication Union (ITU) and the European Telecommunications Standards Institute (ETSI) have released documentation concerning blockchain terminology, use cases, ecosystem, or architectural aspects (see, e.g.,~\cite{ITU1400,etsi2020permissioned}). Other initiatives like the EU Blockchain Observatory, or the standardization of some DLT properties by W3C or Open Timestamps are also active. %A broader view on the standardization of blockchain is collected in. %an open project to accelerate blockchain innovation within the EU, 

As a counterpart, the offered automation and trust are among the most valuable features of blockchain for trading in future mobile networks. In telecommunications, much research work has been ongoing in the area of blockchain-based Internet of Things and beyond 5G~\cite{nguyen2020blockchain}. Blockchain has been widely envisioned to manage network sharing and trading of network slices for vertical applications~\cite{xu2020blockchain}. Accordingly, blockchain may become the basis of a decentralized and transparent platform for multi-party negotiation between the increasing number of stakeholders (e.g., VNF providers, multiple administrative domains) offering end-to-end network slicing. 

Recent architectural O-RAN~\cite{ORANall} evolutions already account for the RAN sharing use case, where multiple operators can configure and control shared resources via remote, open, and newly standardized interfaces. In this context, the introduction of blockchain technologies can automate, accelerate, and secure the definition of the relationship with a well-defined SLA for the trade of resources. The authors of~\cite{xu2021ran} introduced an O-RAN-based architecture to conduct zero-trust mutual authentication over the unique blockchain-enabled routers, and switch functions with the identification exchanged over the blockchain. The main novelty proposed by~\cite{xu2021ran} is to introduce blockchain to decentralize RAN security aspects, in contrast to the current centralized solutions based on authentication and third-party trust. Additionally, other contributions target the vision of blockchain to facilitate resource trading in the RAN segment~\cite{maksymyuk2020blockchain, togou2020dbns}. %In~\cite{maksymyuk2020blockchain}, the authors proposed a protocol for blockchain-based spectrum trading in 5G/6G networks. Finally, the work in~\cite{togou2020dbns} described an architecture that facilitates the dynamic leasing of resources among network operators to support cross-domain services. The cornerstone of this architecture is a brokering layer that relies on a blockchain-based bidding system to exchange resources.

%%%%%%%%%%%%%%%%%%%%%%%%%%%%
%% ARCHITECTURE
%%%%%%%%%%%%%%%%%%%%%%%%%%%%
\section{Blockchain-enabled O-RAN}
\label{section:architecture}
We propose to enrich the O-RAN architecture with new functional blocks for automatic, dynamic sharing of RAN resources through blockchain. The functional architecture is depicted in Fig.~\ref{fig:functionalarchitecture}, where the newly proposed modules are outer lined in red. The introduction of blockchain can automate, accelerate and secure the RAN sharing start-up phase between operator~$i$ ($O_i$) and operator~$j$ ($O_j$) to define the relationship with SLAs for resource trading. Fig.~\ref{fig:functionalarchitecture} represents the simple example of two operators, but multiple parties can take part in this start-up phase, which is facilitated by the introduction of BC. Furthermore, we assume that both operators have the same coverage and service capabilities, and leave open the analysis of the proposed sharing mechanisms under heterogeneous conditions (e.g., different coverage areas, capacities, technologies, frequencies, prices). Specifically, $O_i$ makes its {O-RAN} infrastructure available and hosts the virtual RAN functions of $O_j$ (notice that the antenna resources are shared as well). Remote configuration and monitoring of instantiated VNFs in the host infrastructure of $O_i$ is allowed through O1, O2, and E2 interfaces (defined as entity/remote in the figure). Performance monitoring is implemented during the running phase and complemented by AI functionalities offered by RIC modules, to ensure SLA enforcement. 

The proposed architecture includes different additions to the O-RAN baseline, like the Business Support System (BSS), also present in ETSI-NFV architecture, for the operator to deliver product, customer, and revenue management (billing). The SMO, besides the already introduced non-RT RIC, includes the admission control functionality that decides whether new VNFs can be leased on the site infrastructure. This information is fed to the newly introduced \textit{distributed RAN sharing broker}. In addition, we propose a \textit{marketplace} (potentially managed by a third party) where operators can opt to advertise the infrastructure they are willing to offer, a \textit{BC/SC system}, and the following set of new interfaces: 
\begin{itemize}
    \item ItF-BC1: an SMO internal interface between the admission control and the RAN sharing broker.
    \item ItF-BC2: different broker internal interfaces across its elements.
    \item ItF-BC3: an interface between the BSS and the BC/SC system.
    \item ItF-BC4: an interface between the SMO and the BC/SC system, which can be newly defined or can reuse already available interfaces proposed by O-RAN and 3GPP (i.e., O2, Itf-N, Itf-B, and Type~5 interface).
    \item Vertical industries and OTT providers may also interact with the blockchain and the operators providing infrastructure through the so-called Service Capability Exposure Function (SCEF).
\end{itemize}

The distributed \textit{RAN sharing broker} manages and offers RAN VNFs. The concept of broker has been previously introduced in~\cite{samdanis2016network} for network slicing, as a management point to gather requests from multiple parties such as MVNOs, OTTs, or vertical providers, and to perform admission control. The blocks included in the distributed RAN sharing broker are:
\begin{itemize}
    \item \textbf{Smart contract (SC) creator}: It is in charge of mapping requirements and preferences into a smart contract format so that other operators are informed about the selected service from a marketplace, or about information to participate in an auction. The smart contract defines the SLA that needs to be monitored during the running phase, including information such as the resources type, required QoS, service duration, or tolerance indicators. Once the smart contract is compiled and distributed, the auction engine is triggered (where applicable).
    \item \textbf{Auction engine}: It is in charge of handling the auction process. For the requesting operator, the auction engine collects the bids submitted by the operators, selects the winner based on a precise algorithm for UEs' expected satisfaction, and triggers the communication between the SMOs of the requesting and winner operators. The auction starts after deploying the smart contract, and is concluded after a predefined period (e.g., maximum time, bids number limit, achievement of target price). For candidate operators, the auction engine defines the bid to be submitted based on internal bid algorithms depending on business variables, available resources, resources price, special discounts, and expected return of investment. %During each auction process, bids and decisions are securely recorded in the blockchain.
    \item \textbf{Blockchan (BC) adaptor}: It is the entity in charge of handling the communication with the blockchain and registering the operator to it.
\end{itemize}

\begin{figure}[ht!]
\centering
\includegraphics[width=1\columnwidth]{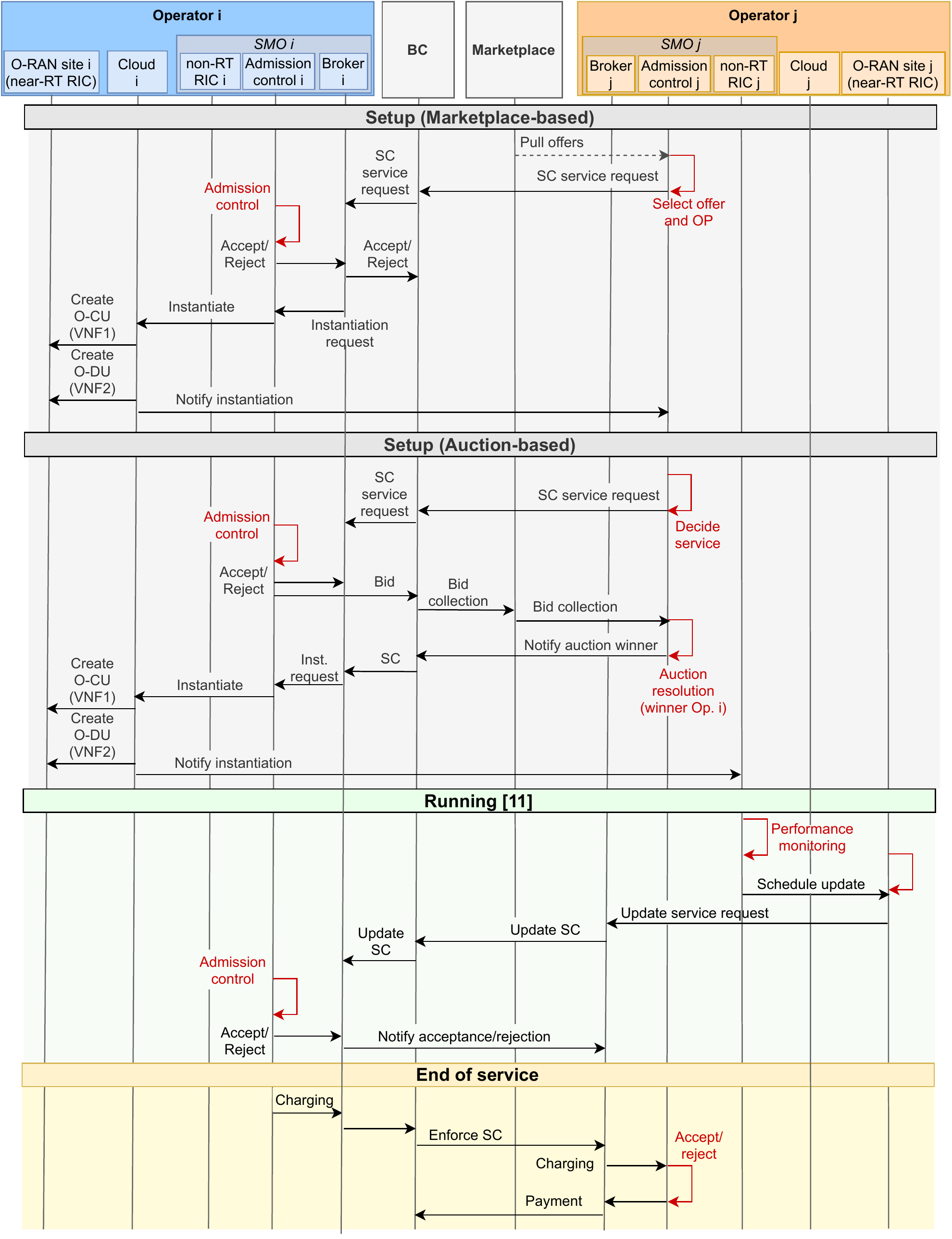}
\caption{Flow diagram of the blockchain-enabled O-RAN functional architecture, for marketplace and auction-oriented cases.}
\label{fig:functionalarchitectureflowdiagram}
\end{figure}

The blockchain in the proposed architecture is private, so only operators deploying a RAN sharing broker can join it. Two mechanisms for RAN sharing are analyzed (see Fig.~\ref{fig:functionalarchitectureflowdiagram}):
\begin{itemize}
    \item \textbf{Marketplace-oriented}: All operators with VNFs, infrastructure, and resources to offer, advertise them with specific prices in a catalog, where other stakeholders select the desired service. To participate in the marketplace, the operator registers the information of interest (e.g., available resources, prices) in the blockchain. In the flow diagram, $O_j$ is interested in leasing a VNF from another operator, so it accesses the marketplace to acquire one. After selecting a sharing provider $O_i$ (based on, e.g., price, availability, coverage area), the smart contract creator in $O_j$'s broker prepares the SLA with the offer from the marketplace. This is distributed in the blockchain and received by $O_i$'s broker, which translates the smart contract into requirements for admission control. If the request is accepted, a VNF is instantiated in $O_i$'s infrastructure to be remotely configured (during the configuration phase) by $O_j$ through the open interfaces defined in O-RAN (O1/O2 remote). During the running phase, the SLA is monitored through RICs at $O_j$ site. Monitoring support from the hosted VNFs, with radio state reports, is realized through the O-RAN native E2-remote interface. Service updates are sent to the broker for preparing new smart contracts, which are registered in the blockchain and evaluated for admission control.
    \item \textbf{Auction-oriented}: In this procedure, the serving $O_i$ is selected using an auction procedure, which is expected to better fit the offer of service and provide a more efficient resource usage. The $O_j$ interested in leasing resources from another operator defines the requirements and desired price for the needed resources and sends them to the smart contract creator of the RAN sharing broker of $O_j$. The smart contract creator prepares a smart contract, which is distributed through the blockchain. Other operators interested in offering to lease their resources, evaluate availability through admission control functionality, and if the resources are available, participate in the auction submitting a bid through their auction engine. The auction engines distribute the bids from the candidate operators through the blockchain. $O_j$'s auction engine collects the bids and decides the target $O_i$ to host the desired VNFs. From this moment, the procedure is similar to the marketplace case. The provisioning request is sent to the hosting operator, to the RAN sharing broker, and an instantiation request is sent to the orchestrator to be instantiated in the platform of $O_i$ site. 
\end{itemize}

The marketplace is expected to boost the speed at which network resources are shared, to allow for fast price comparison and to provide ready-to-go deployable solutions. In contrast, the auction mechanism can provide more personalized services, thus potentially improving efficiency. A further evaluation of the marketplace and the auction solutions is provided in what follows.

%%%%%%%%%%%%%%%%%%%%%%%%%%%%
%% USE CASE
%%%%%%%%%%%%%%%%%%%%%%%%%%%%
\section{Blockchain-enabled RAN sharing: opportunities and challenges}
\label{section:results}
The introduction of blockchain-enabled RAN sharing offers operators new opportunities for new revenue incomes by capitalizing on available infrastructure. In this section, we analyze the different opportunities and challenges, as well as the trade-offs among them, of the proposed blockchain-enabled RAN sharing architectural framework. The most important opportunities offered by Blockchain and smart contracts to traditional RAN sharing are:
\begin{enumerate}
    \item \textbf{Automated management:} By removing long interactions with third parties in the negotiation for sharing resources, improved network management efficiency is expected.
    \item \textbf{Resources efficiency:} Through automated RAN sharing, resources can be fully harnessed by multiple parties, thus leading to a higher resources availability, more coverage, and, consequently, improved UEs' satisfaction.
    \item \textbf{Competitiveness:} With the entry of new actors to the RAN sharing ecosystem, competitiveness is expected to be boosted, which leads to further service diversification and additional possibilities to improve the network infrastructure. This is expected to attract more investments in the network.
    \item \textbf{Auditablity:} The traceability of every interaction in a blockchain allows for improved trust and transparency in RAN sharing. This is an important aspect to take into account, considering the recent issues raised by, e.g., Ericsson warning on O-RAN security~\cite{boswell2020security}.
\end{enumerate}

Concerning the challenges, Blockchain systems have several well-documented drawbacks to take into account when considering their inclusion in an established trading system: 
\begin{enumerate}
    \item \textbf{Communication overhead:} The communication among peer nodes and miners in blockchain generates overhead, which increases with the number of participants and the duration of the requested services (accurate short-term requests vs long-term fixed contracts).
    \item \textbf{Transaction confirmation latency:} The distribution of transactions and blocks across the blockchain determines the delay for instantiating RAN functions in the infrastructure providers' platform. The transaction confirmation delay in a blockchain strongly depends on the adopted consensus mechanism and other parameters like the mining difficulty. A detailed model to estimate the transaction confirmation latency based on those parameters can be found in~\cite{FWilhelmi_PIMRC}.
    \item \textbf{Stability:} The stability of a blockchain is strongly related to the network consensus. In this regard, forks may threaten the blockchain stability. A fork is a split of a blockchain, which occurs when the difference in time between two nodes finding the nonce is lower than the time required by a node to distribute the winning block through the blockchain. As a result, the performance of a blockchain may be jeopardized by a low network capacity. Moreover, game-theoretical aspects may motivate selfish behaviors among miners (e.g., releasing a block at the appropriate time to gain an advantage), thus adding instability to the blockchain. %Forks may incur additional transaction confirmation delay. 
    \item \textbf{Scalability:} Given the nature of blockchain, whereby all the transactions need to be propagated and stored, an increase in the number of blockchain users and transactions can represent both a communication and a storage issue. Besides, blockchain is originally limited by the block size and the mining difficulty (e.g., the case of Bitcoin), which limits the effective transactions rate. 
\end{enumerate}

\begin{figure}[ht!]
\centering
\includegraphics[width=.9\linewidth]{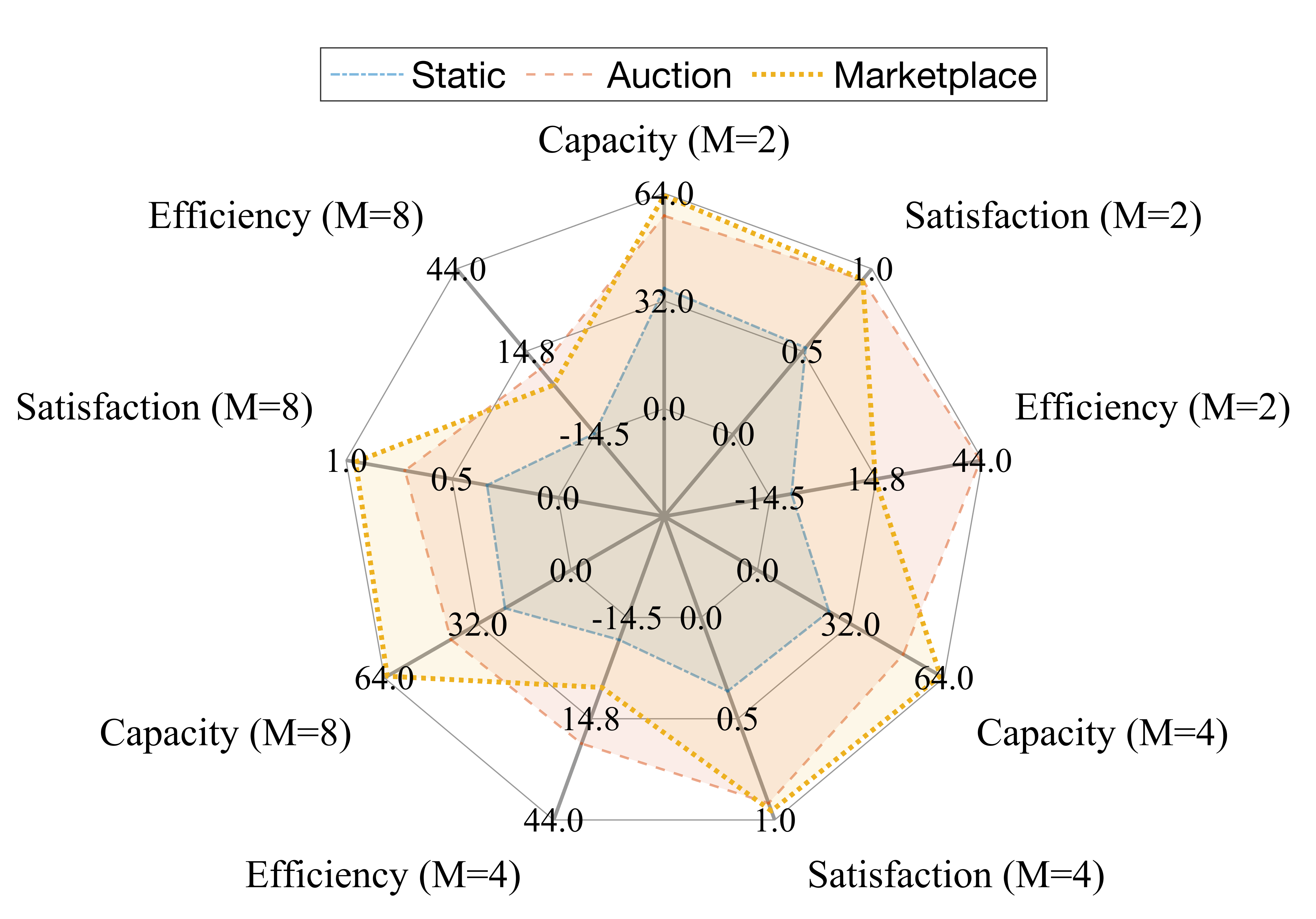}
\caption{Performance achieved by the different blockchain-enabled sharing mechansism compared to the static situation, for M = $\{2, 4, 8\}$ operators.}
\label{fig:performance}
\end{figure}

For performance evaluation, we run simulation campaigns on typical cellular random deployments of up to  up to 19~cells and 200 UEs, where operators owning phyisical infrastructure can virtually split it for RAN sharing.\footnote{All the source code and background information of simulation results are openly available at \textit{\url{https://github.com/fwilhelmi/blockchain_enabled_ran_architecture}, accessed on Oct. 15, 2021}}. Fig.~\ref{fig:performance} compares performance in terms of capacity, UE satisfaction, and efficiency for \textit{(i)} static scenario where RAN sharing is not enabled, \textit{(ii)} marketplace, and \textit{(iii)} auction-oriented approaches. In particular, we compute (1) the UE capacity (in Mbps), as a function of the set of resources allocated from the operator and its signal to noise and interference ratio (SINR); (2) the UE satisfaction (between 0 and 1), as a function of the obtained service and the price paid for it%~\cite{giupponi2008novel}
; (3) the efficiency of each approach (as a ratio), by indicating the degree to which base station resources are appropriately assigned to UEs to meet their needs.

As shown in Fig.~\ref{fig:performance}, both auction and marketplace solutions outperform the static one. The automation provided by blockchain allows improving the utilization of the infrastructure, which leads to increased gains and profits for both sharing and leasing operators. In particular, the marketplace provides higher capacity and satisfaction than the auction approach, since new UEs can be served earlier with the spare capacity already available at the operator side. Regarding efficiency, the auction approach outperforms the marketplace option, because the resources allocated to the UEs can be perfectly adjusted to their demands (a resource sharing contract is done on a per UE basis), instead of mapping to pre-configured offers in the marketplace. The marketplace offers a simpler allocation protocol, but at the price of lower flexibility and less tailored service (sharing requests are only submitted when the available resources are not enough to fulfill UE demands).

\begin{figure}[ht!]
\centering
\includegraphics[width=.85\linewidth]{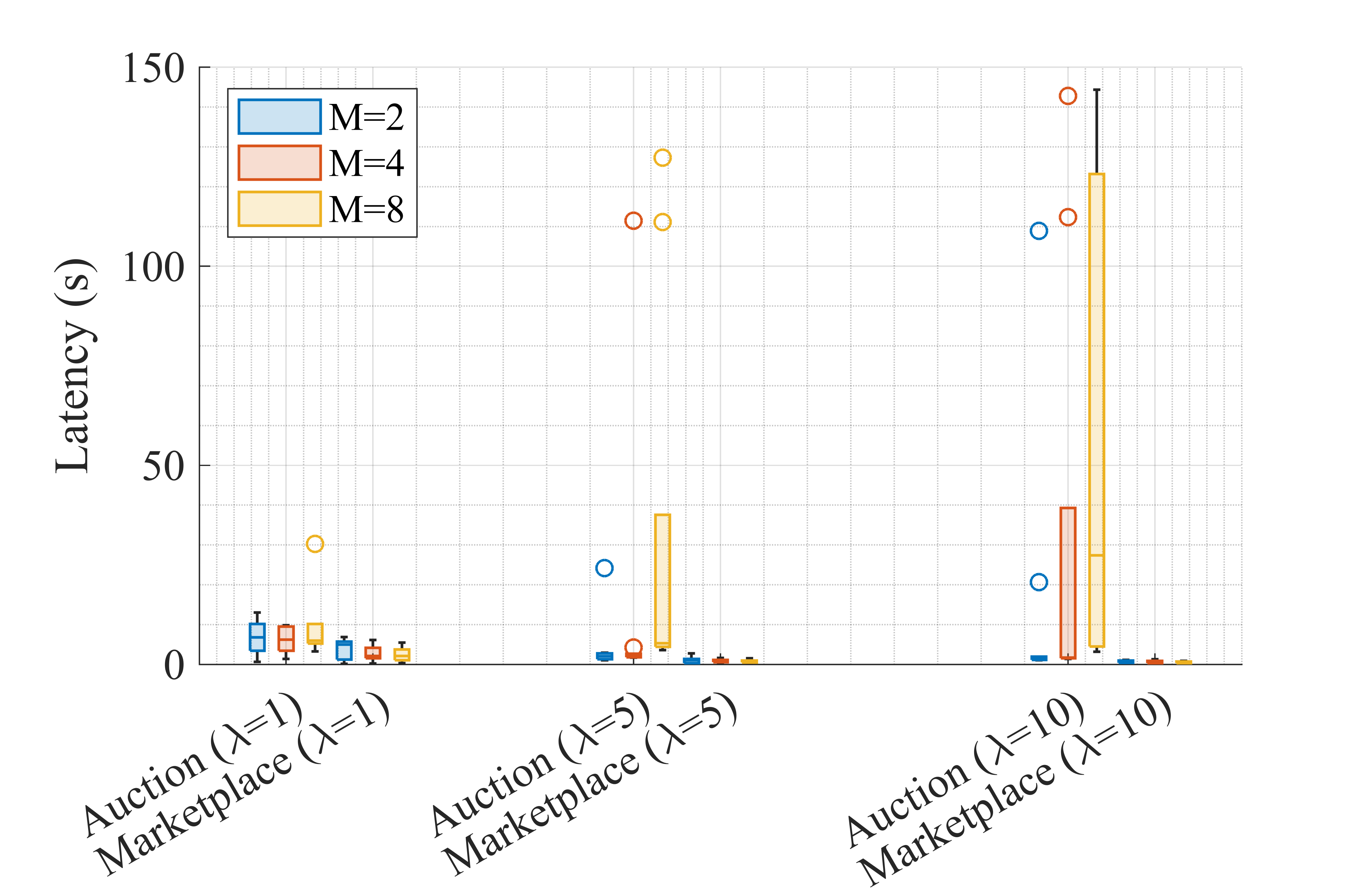}
\caption{Blockchain delay experienced by the auction and marketplace-oriented RAN sharing solutions.}%, different block sizes, user arrival rates, and number of operators.
\label{fig:delay}
\end{figure}

\begin{figure}[ht!]
\centering
\includegraphics[width=.85\linewidth]{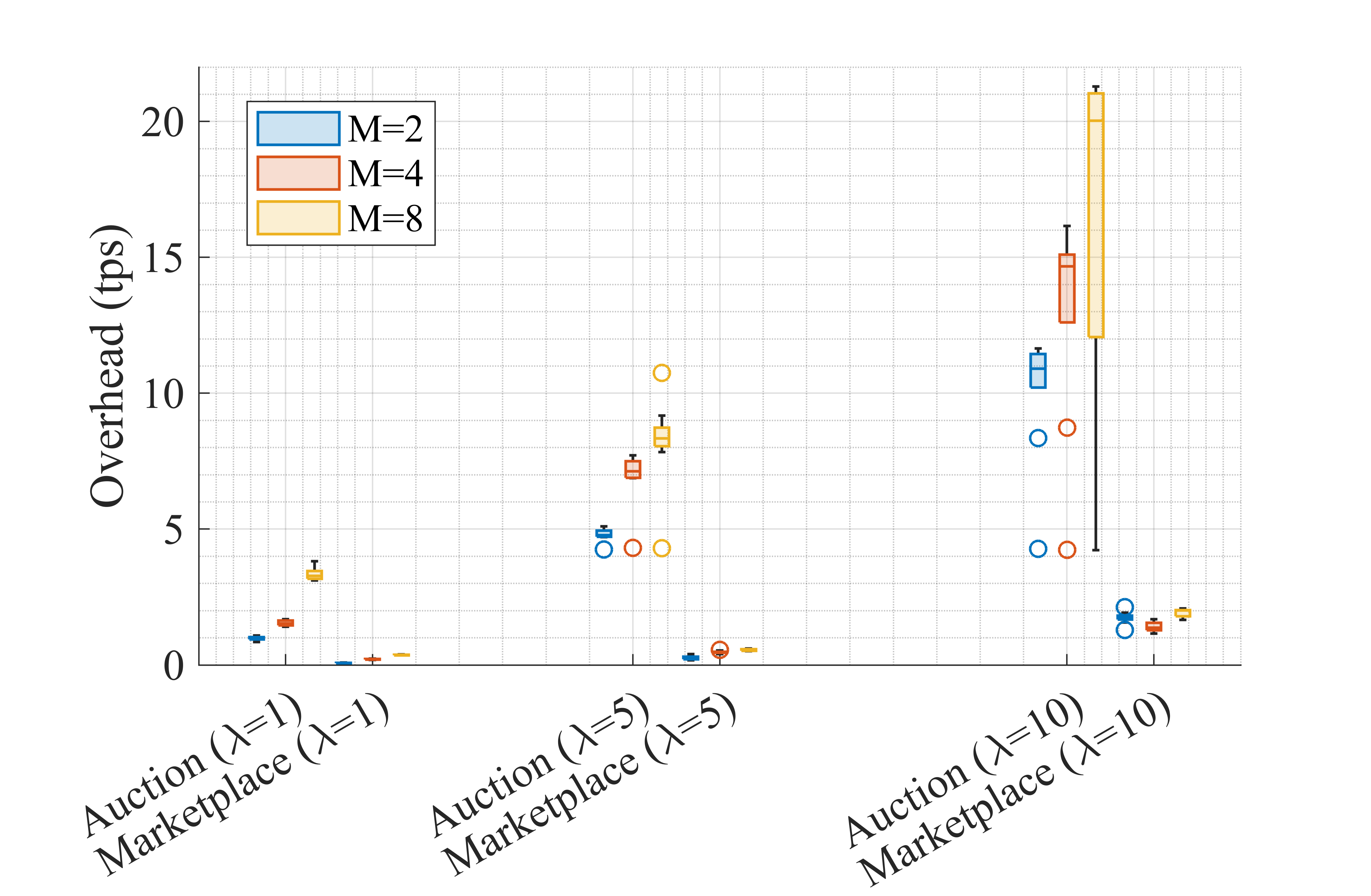}
\caption{Blockchain overhead experienced by the auction and marketplace-oriented RAN sharing solutions.}
\label{fig:overhead}
\end{figure}

We now showcase trade-offs by focusing on the overhead and delay incurred by the proposed RAN sharing solutions. Notice that we consider the total time needed to define the smart contract with the agreed service and to distribute it through the blockchain. This delay is added to the baseline O-RAN delay between the service request and the instantiation of the VNF at a given operator's infrastructure site. In the marketplace case, the blockchain delay includes the time for propagating the service request and automatically enforcing it based on marketplace offers, whereas the auction-based procedure also includes the distribution of bids over the blockchain. Fig.~\ref{fig:delay} and Fig.~\ref{fig:overhead} illustrate the blockchain delay and overhead, respectively, associated to both auction and marketplace-based solutions. The results include different numbers of operators (M = \{2, 4, 8\}), UE request rates ($\lambda$ = \{1, 5, 10\} requests per second), and block size values (from 3,000 to 30,000 bits).

%\begin{figure}[ht!]
%\centering
%\subfigure[]{\includegraphics[width=\columnwidth]{boxplot_combined_delay.png}\label{fig:2}} 
%\subfigure[]{\includegraphics[width=\columnwidth]{boxplot_combined_overhead.png}\label{fig:1}}
%\caption{BC performance for the auction and marketplace-oriented RAN sharing solutions, different block sizes, user arrival rates, and number of operators: a) delay; b) overhead.}
%\label{fig:delay_overhead}
%\end{figure}
We observe that, in spite of the benefits introduced by blockchain, in terms of security and immutability, the auction-based mechanism leads to higher delay and overhead than the marketplace approach. The reason is that the auction leads to a higher number of blockchain transactions, which result from generating tailored service requests and auction bids. This trend is exacerbated as the number of RAN user requests and operators increases. Provided that admission control is properly done, the marketplace option ensures that the delay and overhead do not increase exponentially with the number of operators, which makes it a cost-effective solution for automating RAN sharing procedures in future communications systems. Otherwise, the delay would increase. Moreover, the marketplace delay decreases with the number of RAN user arrivals, which improves the blockchain's efficiency by reducing the overhead (blocks are filled with transactions) and leading to lower waiting times between mined blocks (blocks are generated faster). As for the block size (included in each boxplot), a higher variability is observed in the auction-based approach, which is more susceptible to performance changes for different blockchain parameters. %A more detailed analysis on the impact of blockchain parameters for communications can be found in~\cite{FWilhelmi_PIMRC}.

%%%%%%%%%%%%%%%%%%%%%%%%%%%%
%% CONCLUSIONS
%%%%%%%%%%%%%%%%%%%%%%%%%%%%
\section{Conclusions}
\label{section:conclusions}

The virtualization of network functions provided by 5G is disrupting the way RAN resources are shared. The O-RAN Alliance is contributing to the development of standards on \textit{intelligent, open, virtualized, and fully interoperable mobile networks}, paving the way to a sharing ecosystem for future RAN with providers operating from the cloud. In such an emerging ecosystem, the advertisement of services and automation of administrative negotiations in the form of an open marketplace may become a fundamental piece to trade resources \textit{as-a-service}. To foster trustworthiness, automation, and an economically driven management of the network, we have proposed to extend the baseline O-RAN architecture with the introduction of a programmable trust concept. More specifically, we have envisioned the utilization of blockchain technology, which is expected to provide trust and traceability to the automation of O-RAN management. The novel O-RAN-based blockchain-enabled architecture defines new components and procedures to carry out two proposed automated RAN sharing mechanisms, based on auctioning and open marketplaces. Finally, as a prelude to future research, we have analyzed the main trade-offs offered by the proposed solution through simulation results. We have shown the superiority of the auction and marketplace approaches compared to a static solution in terms of capacity, satisfaction, and efficiency. However, the new overheads and delays introduced by the blockchain need to be carefully considered in the design of future blockchain-enabled RAN architectures. Future works will include technical contributions aimed to optimize operators' selection criteria and to overcome the limitations associated with the introduction of blockchain in O-RAN architectures, as already identified in this work. Furthermore, deeper analyses and proof-of-concept implementations are required to properly understand the costs and benefits associated with blockchain-enabled RAN sharing.

\section*{Acknowledgment}
This work was funded by the IN CERCA grant from the Secretaria d'Universitats i Recerca del departament d'Empresa i Coneixement de la Generalitat de Catalunya, and partially from the Spanish MINECO grant TEC2017-88373-R (5G-REFINE) and Generalitat de Catalunya grant 2017 SGR 1195.
	
\ifCLASSOPTIONcaptionsoff
\newpage
\fi
	
\bibliographystyle{IEEEtran}
\bibliography{bibliography}

\vspace{-.6cm}
% if you will not have a photo at all:
\begin{IEEEbiographynophoto}{Lorenza Giupponi}
(lorenza.giupponi@cttc.es) holds  a  PhD  from  Universitat  Politecnica  de  Catalunya
(2007). She is a Research Director in CTTC, and a member of the CTTC Executive Committee.
\end{IEEEbiographynophoto}
\vspace{-.5cm}
\begin{IEEEbiographynophoto}{Francesc Wilhelmi}
(fwilhelmi@cttc.cat) holds a Ph.D. in Information and Communication Technologies (2020), from Universitat Pompeu Fabra (UPF). He is currently a researcher at CTTC. 
\end{IEEEbiographynophoto}

\end{document}